\documentclass[11pt]{article}
\def\baselinestretch{1.0}
\catcode`\@=11
\@addtoreset{equation}{section}

\def\single {
                \renewcommand{\baselinestretch}{1}
                \large
                \normalsize
                }
\def\ie{\hbox{\it i.e.}}	
	
\def\etal{\hbox{\it et al.}}

\let\vev\VEV

\def\abs#1{\left| #1\right|}

\def\ltap{\mathop{\raisebox{-.4ex}{\rlap{$\sim$}}
\raisebox{.4ex}{$<$}}}

\def\beq{\begin{equation}}
\def\eeq{\end{equation}}

\def\bea{\begin{eqnarray}}
\def\eea{\end{eqnarray}}
\def\underline#1{\relax\ifmmode\@@underline#1\else
	$\@@underline{\hbox{#1}}$\relax\fi}
\@twosidetrue

\catcode`@=12

\catcode`\@=11

\def\section{\@startsection{section}{1}{\z@}{3.5ex plus 1ex minus
   .2ex}{2.3ex plus .2ex}{\bf}}
\def\subsection{\@startsection{subsection}{2}{\z@}{2.0ex plus 1ex minus
   .2ex}{1.0ex plus .2ex}{\em}}
\def\subsubsection{\@startsection{subsubsection}{2}{\z@}{1.25ex plus 1ex minus
   .2ex}{0.5ex plus .2ex}{\em}}
\def\ps@ppt{\def\@oddhead{hep-ph/9511438 \hfil
FERMILAB--CONF--95/353--T}\def\@evenhead{hep-ph/95mmnnn \hfil
FERMILAB--CONF--95/353--T}}
\def\ps@titleppt{
 \def\@oddfoot{\hfill{Opening Lecture at the Second {\it Rencontres du
 Vietnam,} October 21--28,
 1995.}\hfill}%

\def\@oddhead{hep-ph/9511438 \hfil FERMILAB--CONF--95/353--T}}
\catcode`\@=12

\relax
\def\r#1{\ignorespaces $^{\hbox{{\scriptsize #1) }}}$}
\def\figcap{\section*{Figure Captions\markboth
	{FIGURECAPTIONS}{FIGURECAPTIONS}}\list
	{Fig. \arabic{enumi}:\hfill}{\settowidth\labelwidth{Fig. 999:}
	\leftmargin\labelwidth
	\advance\leftmargin\labelsep\usecounter{enumi}}}
 \relax
\def\tablecap{\section*{Table Captions\markboth
	{TABLECAPTIONS}{TABLECAPTIONS}}\list
	{Table \arabic{enumi}:\hfill}{\settowidth\labelwidth{Table 999:}
	\leftmargin\labelwidth
	\advance\leftmargin\labelsep\usecounter{enumi}}}
 \relax
\def\reflist{\section*{Footnotes and References\markboth
	{REFLIST}{REFLIST}}\frenchspacing\list
	{\arabic{enumi}.\hfill}{\settowidth\labelwidth{99.}
	\leftmargin\labelwidth
	\advance\leftmargin\labelsep\usecounter{enumi}}}
 \relax

\input BoxedEPS
\SetRokickiEPSFSpecial
\HideDisplacementBoxes
\newcommand{\eqn}[1]{(\ref{#1})}

\def\gev{\hbox{ GeV}}
\def\tev{\hbox{ TeV}}

\def\tevcc{\hbox{ TeV}\!/\!c^{2}}

\def\sm{$SU(3)_c\otimes SU(2)_L\otimes U(1)_Y$}

\def\cm{\hbox{ cm}}
\def\onetev{1-TeV scale}
\begin{document}
\single
\begin{centering}
{\vphantom{\sc .}}
\vskip4cm
	{\bf PARTICLE PHYSICS: THEMES AND CHALLENGES} \\[12pt]
		Chris Quigg \\
	Fermi National Accelerator Laboratory \\
	P.O. Box 500, Batavia, Illinois 60510 USA \\[3.0cm]
\end{centering}
%\onepointfive
\section{Introduction}
We gather in Ho Chi Minh City with high expectations for the future of
particle physics and high hopes for the future of science in Vietnam.
We are in the midst of a revolution in our perceptions of nature, when
the achievements of our science have brought our insights closer to
everyday life than ever before.  I will devote this lecture to seven
themes that express the essence of our understanding---and our
possibilities.

\section{Elementarity}

One of the pillars of our understanding is the identification of a
set of fundamental constituents, the leptons
\begin{displaymath}
	\begin{array}{ccc}
		\left(
		\begin{array}{c}
			\nu_{e}  \\
			e
		\end{array}
		\right) & \left(\begin{array}{c}
			\nu_{\mu}  \\
			\mu
		\end{array}
		\right) & \left( \begin{array}{c}
			\nu_{\tau}  \\
			\tau
		\end{array}
		\right)
	\end{array},
\end{displaymath}
and the quarks
\begin{displaymath}
	\begin{array}{ccc}
		\left(
		\begin{array}{c}
			u  \\
			d
		\end{array}
		\right) & \left(\begin{array}{c}
			c  \\
			s
		\end{array}
		\right) & \left( \begin{array}{c}
			t  \\
			b
		\end{array}
		\right)
	\end{array},
\end{displaymath}
which have no internal structure, no size, no form factors, and no
excited states---so far as we know.  The quarks are color triplets, so
experience the strong interactions, whereas the leptons, color
singlets, do not.

The charged leptons and the quarks are Dirac particles with
gyromagnetic ratio $g = 2$ (+ the amount induced by
interactions).  The size of the fermions is smaller than the current
limit of experimental resolution characterized by a radius $R \ltap
10^{-17}\cm$.\r{1}  We don't yet know whether the neutrinos are
massive or not.  If neutrinos do have mass, they may be either Dirac
or Majorana particles.\r{2}

All the experimental evidence leads us to conclude that quarks and
leptons are the fundamental (constituent) degrees of freedom at
current energies.  We regard them as elementary.

What if they were
not?  What if quarks and leptons were composite?

Approaching the compositeness scale from low energies, we would
encounter new contact interactions that correspond to the exchange or
rearrangement of constituents.\r{3}  In quark-quark scattering, the
conventional gluon exchange would be supplemented by a contact term of
geometrical size and unknown Lorentz structure.  In $\bar{p}p$
collisions, this new contribution
would lead to an excess (over QCD) of hadron jets at large values of
the transverse energy, where $\bar{q}q \rightarrow \bar{q} q$ is the dominant
elementary reaction.  In general, the angular distribution of the jets
will differ from the standard QCD shape.  If quarks and leptons have
common constituents, a similar excess will be seen in dilepton
production, from the elementary process $\bar{q}q \rightarrow
\ell^{+}\ell^{-}$.  At still higher energies, we expect to see the
effects of excited $q^{*}$ and $\ell^{*}$ states.\r{4}  Finally, at
energies well above the compositeness scale, quarks and leptons would
begin to manifest form factors.

No experimental evidence except history suggests that
quarks and leptons are composite.  However, compositeness might
explain the fermion mass spectrum, the existence of generations, and
the relationship of quarks and leptons.  No composite model has yet
achieved these breakthroughs,\r{5} so the search for compositeness is
a purely experimental exercise.  The discovery of compositeness would
alter our conception of matter in a fundamental way.
\section{Symmetry}
The other essential ingredient in the standard model is the notion
that continuous local symmetries---gauge symmetries---determine the
character of the fundamental interactions.

The simplest, and classic, example is the derivation of quantum
electrodynamics from local phase invariance.  The quantum mechanics
of a free particle is invariant under global changes of phase of the
wave function,
\begin{equation}
	\psi(x) \rightarrow e^{i\theta}\psi(x).
	\label{glph}
\end{equation}
This is the symmetry associated with charge conservation.  Requiring a
theory invariant under \textit{local} changes of phase,
\begin{equation}
	\psi(x) \rightarrow e^{i\theta(x)}\psi(x),
	\label{loph}
\end{equation}
demands the introduction of a massless vector field, identified as
the photon, and leads to a full theory of electrodynamics, QED.\r{6}

The same general strategy can be applied to any continuous symmetry.
That insight links the problem of building theories of the fundamental
interactions to the search for the right symmetries to gauge.  Let us
review the electroweak theory as an example.

The crucial experimental clues for the construction of a gauge theory
of the weak and electromagnetic interactions are the family pattern
embodied in the left-handed weak-isospin doublets of leptons and
quarks and the universal strength of the charged-current weak
interactions.  It is
straightforward to construct the theory, which I will write down for
one generation of leptons, idealizing the neutrinos as massless.

To incorporate $SU(2)_{L}$ weak-isospin symmetry, we define a
left-handed doublet,
\begin{equation}
	\mathsf{L} \equiv \left(
	\begin{array}{c}
		\nu_{L}  \\
		e_{L}
	\end{array}
	\right) = \left(
	\begin{array}{c}
		\frac{1}{2}(1 - \gamma_{5})\nu  \\[4pt]
		\frac{1}{2}(1 - \gamma_{5})e
	\end{array}
	\right),
	\label{Ldef}
\end{equation}
and a right-handed singlet,
\begin{equation}
	\mathsf{R} \equiv e_{R} = {\textstyle \frac{1}{2}}(1 + \gamma_{5})e .
	\label{Rdef}
\end{equation}
To include electromagnetism, we define the weak hypercharge through
$Q = I_{3} + \frac{1}{2}Y$, so that $Y_{L}=-1$, $Y_{R}=-2$.  The
gauge group $SU(2)_{L}\otimes U(1)_{Y}$ allows two coupling
constants, $g$ for the $SU(2)_{L}$ gauge bosons $b^{1}_{\mu},
b^{2}_{\mu}, b^{3}_{\mu}$, and $\frac{1}{2}g^{\prime}$ for the $U(1)$
gauge boson $\mathcal{A}_{\mu}$.  We may write the Lagrangian as
\begin{equation}
	\mathcal{L} = \mathcal{L}_{\mathrm{gauge}} +
	\mathcal{L}_{\mathrm{leptons}},
	\label{ewlag}
\end{equation}
where
	$\mathcal{L}_{\mathrm{gauge}} =
	-\frac{1}{4}F^{\ell}_{\mu\nu}F^{\ell\mu\nu} -
	\frac{1}{4}f_{\mu\nu}f^{\mu\nu}$,
with
$F^{\ell}_{\mu\nu} = \partial_{\nu}b^{\ell}_{\mu} -
\partial_{\mu}b^{\ell}_{\nu} + g
\varepsilon_{jk\ell}b^{j}_{\mu}b^{k}_{\nu}$ and
$f_{\mu\nu} = \partial_{\nu}\mathcal{A}_{\mu} -
\partial_{\mu}\mathcal{A}_{\nu}$,
and the matter term is
\begin{equation}
	\mathcal{L}_{\mathrm{leptons}}  =  \bar{\mathsf{R}}i\gamma^{\mu}
	\left( \partial_{\mu}+ \frac{ig^{\prime}}{2}\mathcal{A}_{\mu}Y
	\right)\mathsf{R}
	  +  \bar{\mathsf{L}}i\gamma^{\mu}
	\left( \partial_{\mu}+ \frac{ig^{\prime}}{2}\mathcal{A}_{\mu}Y +
	\frac{ig}{2} \vec{\tau} \cdot \vec{b}_{\mu}
	\right)\mathsf{L}.
	\label{leplag}
\end{equation}

Explicit mass terms for the gauge bosons or fermions are
inconsistent with the gauge symmetry.  Accordingly, this theory has a
massless neutrino, a massless electron, and four massless electroweak
gauge bosons.  Nature has a massive electron, a massless neutrino,
three massive gauge bosons, and but one massless
electroweak gauge boson, the photon.  The minimal solution to this
mismatch is to hide the gauge symmetry by means of the Higgs
mechanism.  We introduce a complex weak-isospin doublet of scalar
fields,
\begin{equation}
	\phi = \left(
	\begin{array}{c}
		\phi^{+}  \\
		\phi^{0}
	\end{array}
	\right),
	\label{phidef}
\end{equation}
with weak hypercharge $Y_{\phi}=+1$.  Add to the Lagrangian a piece
\begin{equation}
	\mathcal{L}_{\mathrm{scalar}}=(\mathcal{D}^{\mu}\phi)^{\dagger}
	(\mathcal{D}_{\mu}\phi)  - V(\phi^{\dagger}\phi),
	\label{gaugelag}
\end{equation}
where the gauge-covariant derivative is
$\mathcal{D}_{\mu} = \partial_{\mu}+ {\displaystyle
\frac{ig^{\prime}}{2}}\mathcal{A}_{\mu}Y +
	{\displaystyle \frac{ig}{2}} \vec{\tau} \cdot \vec{b}_{\mu}$, and the Higgs
potential is
$V(\phi^{\dagger}\phi) = \mu^{2}(\phi^{\dagger}\phi) +
|\lambda|(\phi^{\dagger}\phi)^{2}$.  We are also free to add the
interactions of the scalars with the fermions,
\begin{equation}
	\mathcal{L}_{\mathrm{Yukawa}} = -G_{e}\left[
	\bar{\mathsf{R}}(\phi^{\dagger}\mathsf{L}) +
	(\bar{\mathsf{L}}\phi)\mathsf{R} \right].
	\label{yuklep}
\end{equation}

If $\mu^{2} < 0$, the vacuum state corresponds to a nonzero value of
the scalar field, which we choose to be
\begin{equation}
	\vev{\phi}_{0} = \left(
	\begin{array}{c}
		0  \\
		v/\sqrt{2}
	\end{array}
	\right) = \left(
	\begin{array}{c}
		0  \\
		(G_{F}\sqrt{8})^{-1/2}
	\end{array}
	\right).
	\label{phivev}
\end{equation}
(The last identification ensures that the theory reproduces the
low-energy charged-current phenomenology.)  The nonzero value of
$\vev{\phi}_{0}$ hides (or breaks) the $SU(2)_{L}$ and $U(1)_{Y}$
symmetries, but preserves a residual invariance under
$U(1)_{\mathrm{EM}}$.  The spectrum of the broken theory consists of a
massless photon $A_{\mu} = \mathcal{A}_{\mu}\cos\theta_{W} +
b^{3}_{\mu}\sin\theta_{W}$, with coupling
$gg^{\prime}/\sqrt{g^{2}+g^{\prime 2}} \equiv e$; charged vector
bosons $W^{\pm}_{\mu} = (b^{1}_{\mu}\mp i b^{2}_{\mu})/\sqrt{2}$, with
$M_{W}^{2}=\pi\alpha/G_{F}\sqrt{2}\sin^{2}\theta_{W}$; a neutral
intermediate boson $Z_{\mu}=b^{3}_{\mu}\cos\theta_{W} -
\mathcal{A}_{\mu}\sin\theta_{W}$, with $M_{Z} = M_{W}\!/\!\cos\theta_{W}$;
a neutral Higgs scalar, with $M_{H}^{2} = -2\mu^{2} > 0$; and an
electron with mass $m_{e}=G_{e}v/\sqrt{2}$.  The predicted
masses for the $W^{\pm}$ and $Z^{0}$
are expressed in terms of the weak mixing parameter
$\sin^{2}\theta_{W}$, which is measured in neutral-current
reactions.  It is both a triumph and a frustration of the electroweak
theory that spontaneous symmetry breaking plus Yukawa couplings
generates fermion masses, for the Yukawa couplings are not calculable
within the theory.

\section{Consistency}
The leptonic and hadronic charged weak currents are identical in
form, characterized by the left-handed doublets
\begin{equation}
	\begin{array}{cc}
		\left(
		\begin{array}{c}
			\nu_{e}  \\
			e
		\end{array}
		\right)_{\!L}, & 		\left(
		\begin{array}{c}
			u  \\
			d_{\theta}
		\end{array}
		\right)_{\!L},
	\end{array}
	\label{chcu}
\end{equation}
etc.  Renormalizability requires the absence of anomalies---quantum
violations of classical symmetries or conservation laws.  For the
electroweak theory, the condition for anomaly freedom can be expressed
in the requirement
\begin{equation}
	\Delta Q = Q_{\mathrm{R}}-Q_{\mathrm{L}} =
	\sum_{\parbox[b]{40pt}{\tiny
	RH doublets}}\!\!\!\!Q - \!\!\sum_{\parbox[b]{40pt}{\tiny
	LH doublets}}\!\!\!\!Q = 0.
	\label{anom}
\end{equation}
In an electroweak theory based on the lepton doublet $(\nu_{e}\;
e)_{L}$, $\Delta Q^{\mathrm{(leptons)}}=-Q_{\mathrm{L}} = +1 \ne 0$.
To cancel the lepton anomaly, we could add right-handed fermions with
appropriate charges, but no right-handed charged-current interactions
are known.  More to the point, we can add a color triplet of
left-handed quark doublets $(u\; d_{\theta})_{L}$, for which $\Delta
Q^{\mathrm{(quarks)}}= -3(\frac{2}{3} - \frac{1}{3}) = -1$, so that
$\Delta Q = \Delta Q^{\mathrm{(leptons)}} + \Delta
Q^{\mathrm{(quarks)}}=0$.

It is remarkable that a consistent theory of weak and electromagnetic
interactions requires quarks as well as leptons.  This suggests a
deep connection between quarks and leptons that I take as an important
clue toward a more complete theory.

\section{Unity}
Making connections is the essence of scientific progress.  For
monumental examples, think of the amalgamation of electricity and
magnetism and light; of the recognition that heat is atoms in motion,
which brought together thermodynamics and Newtonian mechanics; and of
the realization that the chemical properties of substances are
determined by the atomic and molecular structure of matter.  Each of
these unifications brought new understanding and illuminated
phenomena beyond those that served as motivation.

What progress might we achieve by unifying the quarks and leptons, or
the strong, weak, and electromagnetic interactions described by the
\sm\ gauge theories, or both?  The link between quarks and leptons
implied by anomaly cancellation is reinforced by the following set of
questions: Can we understand why (i) electric charge is quantized?
(ii) $Q_{p}+Q_{e}=0$?  (iii) $Q_{\nu}-Q_{e}=Q_{u}-Q_{d}$?  (iv)
$Q_{d}=Q_{e}/3$?  (v) $Q_{\nu}+Q_{e}+3Q_{u}+3Q_{d}=0$?

What observations motivate a unified theory of the fundamental
interactions?  Beyond the similarities and links between quarks and
leptons, we recognize that the $SU(2)_{L}\otimes U(1)_{Y}$ elctroweak
theory achieves only a partial unification of the weak and
electromagnetic interactions, as evidenced by the fact that
$\sin^{2}\theta_{W}$ is a free parameter of the theory.  Taken
together, quantum chromodynamics and the electroweak theory have
three distinct coupling parameters, $(\alpha_{s},
\alpha_{\mathrm{EM}}, \sin^{2}\theta_{W})$ or, equivalently,
$(\alpha_{3}, \alpha_{2}, \alpha_{1})$.  Might we reduce the number of
independent couplings to two or one?  As we shall review presently,
the evolution of the gauge couplings suggests that coupling-constant
unification might be possible.

The minimal example of a theory that unifies the quarks and leptons
and the fundamental interactions is based on the gauge group
$SU(5)$.\r{7}
The gauge bosons of $SU(5)$ lie in the adjoint $\mathbf{24}$
representation.  Decomposing these particles according to their
$(SU(3)_{c}, SU(2)_{L})_{Y}$ quantum numbers, we recognize
\begin{equation}
	\begin{array}{ccc}
		(\mathbf{8},\mathbf{1})_{0} & : & \mathrm{gluons},  \\
		(\mathbf{1},\mathbf{3})_{0} & : & W^{+}, W^{-}, W_{3},  \\
		(\mathbf{1},\mathbf{1})_{0} & : & \mathcal{A},
	\end{array}
	\label{su5sm}
\end{equation}
the twelve gauge bosons of (unbroken) \sm, plus twelve new force
particles whose existence is implied by the unification:
\begin{equation}
	\begin{array}{ccc}
		(\mathbf{3},\mathbf{2})_{-5/3} & : & X^{-4/3}, Y^{-1/3},  \\
		(\mathbf{3^{*}},\mathbf{2})_{5/3} & : & X^{4/3}, Y^{1/3}.
	\end{array}
	\label{su5lq}
\end{equation}
These additional interactions mediate baryon- and
lepton-number-violating processes.  The fundamental fermions fit in
the $\mathbf{5^{*}}$ and $\mathbf{10}$ representations of $SU(5)$,
with
	$\nu_{e}, e_{L}, d^{c}_{L} \in \mathbf{5^{*}}$ and
	$e^{c}_{L}, u_{L}, d_{L}, u^{c}_{L} \in \mathbf{10}$,
where I have used charge-conjugate fields to represent the
right-handed degrees of freedom.

\begin{figure}[t!]
	\centerline{\BoxedEPSF{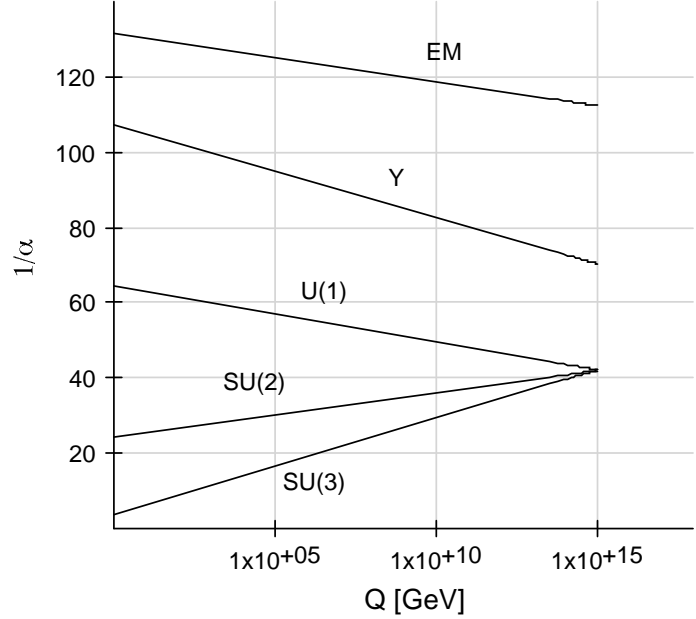 scaled 750}}
	\caption{Evolution of running coupling constants in leading
	logarithmic approximation in the three-generation $SU(5)$ model.}
	\protect\label{evol}
\end{figure}

It is a straightforward matter to compute the evolution of the \sm\
gauge couplings.\r{8}  Writing $\alpha_{i}=g_{i}^{2}/4\pi$, we have to
leading logarithmic approximation
\begin{equation}
	1/\alpha_{3}(Q^{2}) = 1/\alpha_{3}(\mu^{2})+b_{3}\ln(Q^{2}/\mu^{2}),
	\label{run3}
\end{equation}
where
	$4\pi b_{3} = 11 - 2n_{f}/3 = 11 - 4n_{g}/3$,
and $n_{f}(n_{g})$ is the number of active flavors (generations) with
$\hbox{mass} < \sqrt{Q^{2}}$;
\begin{equation}
	1/\alpha_{2}(Q^{2}) = 1/\alpha_{2}(\mu^{2})+b_{2}\ln(Q^{2}/\mu^{2}),
	\label{run2}
\end{equation}
where
	$4\pi b_{2} = (22 - 4n_{g}-{\textstyle \frac{1}{2}})/3$;
\begin{equation}
	(5/3)(1/\alpha_{1}(Q^{2})) = 1/\alpha_{Y}(Q^{2})=
	1/\alpha_{Y}(\mu^{2})+b_{Y}\ln(Q^{2}/\mu^{2}),
	\label{run1}
\end{equation}
where
	$4\pi b_{Y} = -20n_{g}/9$.
We equate $\alpha_{3}, \alpha_{2}, \alpha_{1}$ at the unification
scale $M_{U}$.  To estimate the unification scale, we take
\begin{equation}
	1/\alpha(M_{Z}^{2}) = 1/\alpha_{Y}(M_{Z}^{2}) +
	1/\alpha_{2}(M_{Z}^{2}) = 129.08
	\label{runalpha}
\end{equation}
and
\begin{equation}
	\alpha_{3}(M_{Z}^{2}) = 0.116,
	\label{a3mz}
\end{equation}
whereupon
\begin{equation}
	M_{U} \approx 10^{15}\gev.
	\label{uni}
\end{equation}
The characteristic evolution of the coupling constants is shown in
Figure \ref{evol}.  Reality seems a little different.\r{9}

\begin{figure}[t!]
	\centerline{\BoxedEPSF{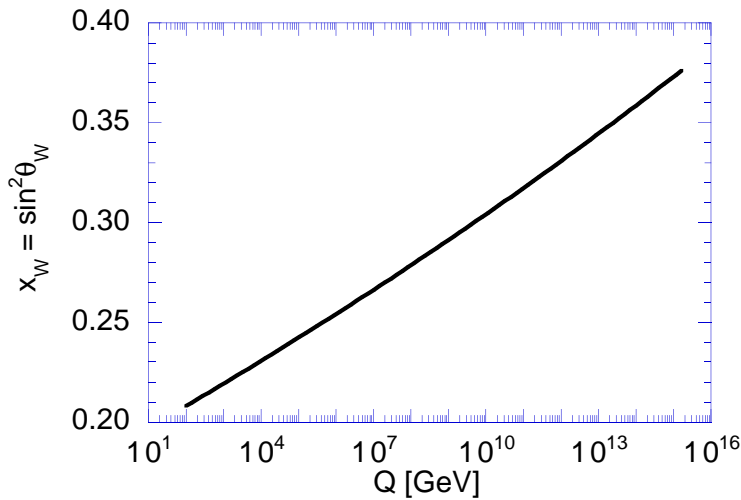 scaled 800}}
	\caption{Evolution of the weak mixing parameter in the
	three-generation $SU(5)$ model.}
	\protect\label{xwfig}
\end{figure}

Of special interest is the evolution of the weak mixing parameter
\begin{equation}
	x_{W}= \sin^{2}\theta_{W} = \alpha/\alpha_{2}.
	\label{xdef}
\end{equation}
The evolution equations for the gauge couplings yield
\begin{eqnarray}
	x_{W}(Q^{2}) &  = & \frac{3}{8} - \alpha(Q^{2})
	\frac{(3b_{Y}-5b_{2})}{8}\ln(Q^{2}/M_{U}^{2}) \\
	 & = & \frac{3}{8} + \frac{55\alpha(Q^{2})}{48\pi}\ln(Q^{2}/M_{U}^{2}),
	\label{xwev}
\end{eqnarray}
which is sketched in Figure \ref{xwfig}.
  The $SU(5)$ prediction at the weak scale is
\begin{equation}
	\left.x_{W}(M_{Z}^{2})\right|_{SU(5)} \approx 0.21,
	\label{xwmz}
\end{equation}
which is both tantalizingly close to, and frustratingly far from, the
LEP--SLD average value\r{10}
\begin{equation}
	\left.x_{W}(M_{Z}^{2})\right|_{\mathrm{LEP+SLD}} = 0.23143 \pm
	0.00028.
	\label{xwlep}
\end{equation}

Let us summarize the standing of the $SU(5)$ example of a unified
theory.  $SU(5)$ contains the standard-model gauge group \sm\ in a
simple group, and thus reduces the number of independent couplings
from three to one.  By construction, the theory gives a correct
description of the charged-current weak interactions.  So far as the
neutral-current interactions are concerned, it predicts a value of
the weak mixing parameter that is close to, but not identical with,
the observed value.  $SU(5)$ naturally explains the quantization of
electric charge.  Since the electric-charge operator $Q$ is a
generator of $SU(5)$, the sum of electric charges over any
representation must be zero.  This means in particular that
$Q(d^{c})=(-1/3)Q(e)$.  Proton decay is possible, through the
action of the color-triplet gauge bosons $X$ and $Y$.\r{11}  And of
course, aspirations remain, even beyond a unified theory of the
strong, weak, and electromagnetic interactions, for gravitation is
omitted from the theory.

\section{Identity}
What makes a bottom quark a bottom quark, or an electron an
electron?  One of the great unsolved problems of the standard model is
how to calculate fermion masses and mixing angles.  In the electroweak
theory, the Higgs mechanism produces fermion masses, as a result of
spontaneous symmetry breaking.  Recall that for a single generation
of leptons, the Yukawa interaction is
\begin{equation}
	\mathcal{L}_{\mathrm{Yukawa}} =
	-G_{e}\left[\bar{\mathsf{R}}(\phi^{\dagger}\mathsf{L}) +
	(\bar{\mathsf{L}}\phi)\mathsf{R}\right],
	\label{lepyuk}
\end{equation}
where the left-handed lepton doublet is
\begin{equation}
	\mathsf{L} = \left(
	\begin{array}{c}
		\nu_{e}  \\
		e
	\end{array}
	\right)_{L}
	\label{Llepdef}
\end{equation}
and $\mathsf{R}=e_{R}$ is the right-handed electron.  The
interaction (\ref{lepyuk}) is the most general Lorentz scalar invariant
under local $SU(3)_{c}\otimes U(1)_{Y}$ transformations.  The
electroweak theory offers no guidance about the value of the Yukawa
coupling.  For the electron, $G_{e} \approx 3 \times 10^{-6}$, while
the analogous coupling for the top quark is $G_{t} \approx 1$.

For three generations of quarks and leptons, we can generalize
(\ref{lepyuk}) to
\begin{equation}
	\mathcal{L}_{\mathrm{Yukawa}} = \bar{u}^{i}_{R}U_{ij}
	(\bar{\phi}^{\dagger}_{u}\mathsf{Q}_{j}) +
	\bar{d}^{i}_{R}D_{ij}(\phi^{\dagger}_{d}\mathsf{Q}_{j}) +
	\bar{e}^{i}_{R}E_{ij}(\phi^{\dagger}_{d}\mathsf{L}_{j}),
	\label{3genyuk}
\end{equation}
where
\begin{equation}
	\begin{array}{cc}
		\mathsf{Q}_{j} = \left(
		\begin{array}{c}
			u_{j}  \\
			d_{j}
		\end{array}
		\right)_{L} & \mathsf{L}_{j} = \left(
		\begin{array}{c}
			\nu_{j}  \\
			\ell_{j}
		\end{array}
		\right)_{L},
	\end{array}
	\label{dubs}
\end{equation}
and $U_{ij}, D_{ij},\hbox{ and }E_{ij}$ are complex $3 \times 3$
matrices.  In the electroweak theory, $\phi_{u}=\phi_{d}=\phi$, with
$\vev{\phi}_{0}$ given by \eqn{phivev}, whereas in the minimal supersymmetric
generalization, $\phi_{u}$ and $\phi_{d}$ are distinct.  The ratio of
their vacuum expectation values is parametrized as $\tan\beta =
v_{u}/v_{d}$.

Unified theories imply relations among fermion masses.  In the example
of $SU(5)$, spontaneous symmetry breaking occurs in two steps.  First,
an adjoint $\mathbf{24}$ of scalars breaks
\begin{equation}
	SU(5) \rightarrow SU(3)_{c}\otimes SU(2)_{L}\otimes U(1)_{Y},
	\label{bk24}
\end{equation}
and gives very large masses to the $X^{\pm 4/3}$ and $Y^{\pm 1/3}$
gauge bosons.  But because the $\mathbf{24}$ does not occur in the
$\bar{\mathsf{L}}\mathsf{R}$ products
	$\mathbf{5^{*}}\otimes\mathbf{10} = \mathbf{5} \oplus
	\mathbf{45}$ and
	$\mathbf{10}\otimes\mathbf{10} = \mathbf{5^{*}} \oplus
	\mathbf{45^{*}} \oplus \mathbf{50^{*}}$,
no fermion masses are generated at this stage.  Electroweak symmetry
is broken by the $SU(5)$ generalization of the usual Higgs mechanism,
a $\mathbf{5}$ of scalars that contains the standard-model Higgs
doublet:
\begin{equation}
	SU(3)_{c}\otimes SU(2)_{L}\otimes U(1)_{Y} \rightarrow
	SU(3)_{c}\otimes U(1)_{\mathrm{EM}}.
	\label{bk5}
\end{equation}
This pattern of symmetry breaking leads to the relations
\begin{equation}
	\begin{array}{ccc}
		m_{e} = m_{d},  &
		m_{\mu} = m_{s}, &
		m_{\tau} = m_{b},
	\end{array}
	\label{eqmass}
\end{equation}
at the unification scale.  The masses of the charge-2/3 quarks are
separate parameters $m_{u}, m_{c}, m_{t}$.

Like the values of coupling constants, the values of particle masses
depend on the scale on which they are observed.  In leading
logarithmic approximation, the running masses of the up-like quarks
evolve as
\begin{eqnarray}
	\ln m_{u,c,t}(\mu) & \approx & \ln m_{u,c,t}(M_{U}) +
	\frac{12}{33-2n_{f}} \ln \left( \frac{\alpha_{3}(\mu)}{\alpha_{U}}\right)
	\nonumber  \\
	 & + & \frac{27}{88 - 8n_{f}}\ln \left(
	 \frac{\alpha_{2}(\mu)}{\alpha_{U}}\right) - \frac{3}{10n_{f}}\ln \left(
	 \frac{\alpha_{1}(\mu)}{\alpha_{U}}\right);
 	\label{upev}
\end{eqnarray}
the down-like quark masses run as
\begin{eqnarray}
	\ln m_{d,s,b}(\mu) & \approx & \ln m_{d,s,b}(M_{U}) +
	\frac{12}{33-2n_{f}} \ln \left( \frac{\alpha_{3}(\mu)}{\alpha_{U}}\right)
	\nonumber  \\
	 & + & \frac{27}{88 - 8n_{f}}\ln \left(
	 \frac{\alpha_{2}(\mu)}{\alpha_{U}}\right) + \frac{3}{20n_{f}}\ln \left(
	 \frac{\alpha_{1}(\mu)}{\alpha_{U}}\right);
 	\label{dnev}
\end{eqnarray}
and the masses of the charged leptons evolve as
\begin{eqnarray}
	\ln m_{e,\mu,\tau}(\mu) & \approx & \ln m_{e,\mu,\tau}(M_{U})
	\nonumber  \\
	 & + & \frac{27}{88 - 8n_{f}}\ln \left(
	 \frac{\alpha_{2}(\mu)}{\alpha_{U}}\right) - \frac{27}{20n_{f}}\ln \left(
	 \frac{\alpha_{1}(\mu)}{\alpha_{U}}\right).
 	\label{lepev}
\end{eqnarray}
Accordingly, the masses of the $b$-quark and the $\tau$-lepton evolve
to different values at low energies:
\begin{equation}
	\ln \left[ \frac{m_{b}(\mu)}{m_{\tau}(\mu)}\right]  \approx
	\frac{12}{33-2n_{f}} \ln \left( \frac{\alpha_{3}(\mu)}{\alpha_{U}}\right)
+ \frac{3}{2n_{f}}\ln \left(
	 \frac{\alpha_{1}(\mu)}{\alpha_{U}}\right).
 	\label{btau}
\end{equation}
Choosing $n_{f}=6$, $1/\alpha_{U}=40$, $1/\alpha_{3}(\mu)=5$, and
$1/\alpha_{1}=65$, we compute (which is to say, predict)
\begin{equation}
	m_{b} = 2.91 m_{\tau} = 5.16\gev\!/\!c^{2},
	\label{btaunum}
\end{equation}
in good agreement with the facts.  To make this simple estimate, I
have neglected the change in evolution at top threshold.  Higgs-boson
contributions, omitted here, are important for the evolution of
heavy-fermion masses.  The top Yukawa coupling plays a crucial role
in supersymmetric models.

Starting from the equalities $m_{s}(M_{U}) = m_{\mu}(M_{U})$ and
$m_{d}(M_{U}) = m_{e}(M_{U})$, equations \eqn{dnev} and \eqn{lepev}
lead to the prediction that at $\mu \approx 1\gev$,
\begin{equation}
	\frac{m_{s}}{m_{d}} = \frac{m_{\mu}}{m_{e}} .
	\label{sdmue}
\end{equation}
This is less successful; empirically, the left-hand-side is about 20,
and the right-hand-side about 200.  A more elaborate scheme for
breaking the electroweak symmetry---say, adding a $\mathbf{45}$ of
scalars---can give rise to a different simple pattern of masses at
the unification scale.  The simple pattern
\begin{equation}
	\begin{array}{cc}
		m_{s}(M_{U}) = \frac{1}{3}m_{\mu}(M_{U}), &
		m_{d}(M_{U}) = 3m_{e}(M_{U})
	\end{array}
	\label{altpatt}
\end{equation}
leads to
\begin{equation}
	\begin{array}{cc}
		m_{s} \approx \frac{4}{3}m_{\mu}, &
		m_{d} \approx 12m_{e},
	\end{array}
	\label{altmass}
\end{equation}
at $\mu = 1\gev$.

The important point in these exercises is not that $SU(5)$ gives us an
understanding of the pattern of fermion masses, but the more general
lesson that a simple pattern at the unification scale can manifest
itself in a complicated (irrational!) pattern at low energies.  This
insight has spawned a new strategy for making sense of the pattern of
fermion masses---and a new industry for theorists.\r{12}  Begin with a
promising unified theory, like supersymmetric $SU(5)$, which has
advantages over ordinary $SU(5)$ for $\sin^{2}\theta_{W}$, coupling
constant unification, and the proton lifetime, or supersymmetric
$SO(10)$, which can accommodate massive neutrinos.  Then find
``textures,'' simple patterns of Yukawa matrices that lead to
successful predictions for masses and mixing angles.  Interpret these
in terms of patterns of electroweak symmetry breaking.  Finally, seek
a derivation of---or at least a motivating principle for---the winning
entry.  The proof that this program has predictive power is that some
schemes fail for $m_{t}$ or $|V_{cb}|$.

\section{Opportunity}
As successful as the electroweak theory is in describing experimental
observations,\r{13} we do not need hints from experiment to know that
the theory is incomplete.\r{14}  We have only to look at the many
parameters of the \sm\ gauge theories of the strong, weak, and
electromagnetic interactions to see opportunities for a more
predictive theory.  The 6 quark masses, 3 charged-lepton masses, 4
quark-mixing parameters, 3 coupling constants, 2 parameters of the
Higgs potential, and 1 strong (\textit{CP}) phase make 19 parameters whose
values are not explained by the standard model.  Seventeen of these
numbers lie in the domain of the electroweak theory.  Next, we can
inquire into the self-consistency and naturalness of the electroweak
theory.  The hierarchy, naturalness, and triviality problems indicate
that the electroweak theory is not complete.

As an illustration of these shortcomings, let us ask why the
electroweak scale is small.  Note that we do have some understanding,
from the evolution of coupling constants down from the unification
scale, of why the strong interaction becomes strong at a scale of
about $1\gev$.

The $SU(2)_L \otimes U(1)_Y$ electroweak theory does not explain how the
scale of electroweak symmetry breaking is maintained in the presence
of quantum corrections.  The problem of the scalar sector can be
summarized neatly as follows.\r{15}  The Higgs potential is
\begin{equation}
      V(\phi^\dagger \phi) = \mu^2(\phi^\dagger \phi) +
\abs{\lambda}(\phi^\dagger \phi)^2 \;.
\end{equation}
With $\mu^2$ chosen to be less than zero, the electroweak symmetry is
spontaneously broken down to the $U(1)$ of electromagnetism, as the
scalar field acquires a vacuum expectation value that is fixed by the
low-energy
phenomenology,
\begin{equation}
	\vev{\phi}_0 = \left( \begin{array}{c}
	0 \\ \sqrt{-\mu^2/2|\lambda|} \end{array}\right) \equiv
	\left( \begin{array}{c}
	     0 \\ v/\sqrt{2}
	\end{array}\right) = (G_F\sqrt 8)^{-1/2}
		\approx 175 {\rm \;GeV}\;.
\end{equation}
Three of the scalar degrees of freedom become the longitudinal
components of the intermediate vector bosons $W^{+}, W^{-},\hbox{ and
}Z^{0}$.  The fourth emerges as a massive scalar particle, the Higgs
boson, with mass given by
\begin{equation}
	M_{H}^{2} = 2 |\lambda| v^{2} .
	\label{Hmass}
\end{equation}

Beyond the classical approximation, scalar mass parameters receive
quantum corrections from loops that contain particles of spins
$J=1, 1/2$, and $0$:
\begin{equation}
\BoxedEPSF{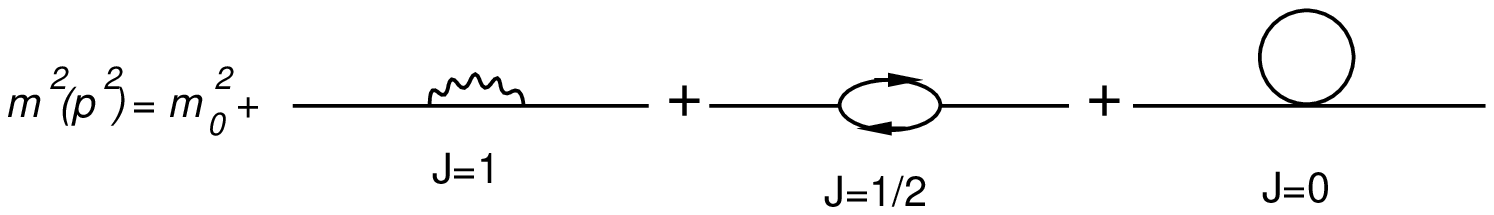 scaled 800}
\label{loopy}
\end{equation}
The loop integrals are potentially divergent.  Symbolically, we may
summarize the content of \eqn{loopy} as
\begin{equation}
	m^2(p^2) = m^2(\Lambda^2) + Cg^2\int^{\Lambda^2}_{p^2}dk^2
	+ \cdots \;, \label{massevol}
\end{equation}
where $\Lambda$ defines a reference scale at which the value of
$m^2$ is known, $g$ is the coupling constant of the theory, and the
coefficient $C$ is calculable in any particular theory.
Instead of dealing with the relationship between observables and
parameters of the Lagrangian, we choose to describe the variation of
an observable with the momentum scale.  In order for the mass shifts
induced by radiative corrections to remain under control (\ie , not to
greatly exceed the value measured on the laboratory scale), either
$\Lambda$ must be small, so the range of integration is not
enormous, or new physics must intervene to cut off the integral.

If the fundamental interactions are described by an
$SU(3)_c\otimes SU(2)_L\otimes U(1)_Y$ gauge symmetry, \ie, by quantum
chromodynamics and the electroweak theory, then the
natural reference scale is the Planck mass,
\begin{equation}
	\Lambda \sim M_{\rm Planck} \approx 10^{19} {\rm \; GeV}\;.
\end{equation}
In a unified theory of the strong, weak, and electromagnetic
interactions, the natural scale is the unification scale,
\begin{equation}
	\Lambda \sim M_U \approx 10^{15}\hbox{-}10^{16} {\rm \; GeV}\;.
\end{equation}
Both estimates are very large compared to the scale of electroweak
symmetry breaking.  We are therefore assured that new physics must
intervene at an energy of approximately $1\tev$, in order that the
shifts in $m$ not be much larger than $v\!/\!\sqrt{2}$.

Only a few distinct scenarios for controlling the
contribution of the integral in \eqn{massevol} can be envisaged.  The
supersymmetric solution\r{16} is especially elegant.  Exploiting the fact
that fermion loops contribute with an overall minus sign (because of
Fermi statistics), supersymmetry balances the contributions of fermion
and boson loops.  In the limit of unbroken supersymmetry, in which the
masses of bosons are degenerate with those of their fermion
counterparts, the cancellation is exact:
\begin{equation}
	\sum_{{i={\rm fermions \atop + bosons}}}C_i\int dk^2 = 0\;.
\end{equation}
If the supersymmetry is broken (as it must be in our world), the
contribution of the integrals may still be acceptably small if the
fermion-boson mass splittings $\Delta M$ are not too large.  The
condition that $g^2\Delta M^2$ be ``small enough'' leads to the
requirement that superpartner masses be less than about
$1\tevcc$.

A second solution to the problem of the enormous range of integration in
\eqn{massevol} is offered by theories of dynamical symmetry breaking such as
technicolor.\r{17} In technicolor models, the Higgs boson is composite, and
new physics arises on the scale of its binding, $\Lambda_{TC} \simeq
O(1~{\rm TeV})$. Thus the effective range of integration is cut off, and
mass shifts are under control.

A third possibility is that the gauge sector becomes strongly
interacting.\r{18} This would give rise to $WW$ resonances, multiple
production of gauge bosons, and other new phenomena at energies of
$1\tev$  or so.  It is likely that a scalar bound state---a quasi-Higgs
boson---would emerge with a mass less than about $1\tevcc$.

We cannot avoid the conclusion that some new physics must occur on
the \onetev.  This is the principal sharp motivation for multi-TeV
hadron colliders---for the LHC.  We seek to complete our understanding
of electroweak symmetry breaking.  A thorough investigation of the
\onetev\ promises to solve the problem of gauge-boson masses and give
us insight, if not a complete solution, into the origin of fermion
masses.  The large step in energy and sensitivity will also test the
underpinnings of the standard model by allowing us to search for new
forces, for composite quarks and leptons, and for new forms of matter.

\section{Relevance}
Physics is possible because we can analyze phenomena on one energy
scale without understanding all energy scales.  In other words, we
need not understand everything before we can begin to answer
something.  In quantum field theory, it is frequently possible to
identify the relevant degrees of freedom on some energy scale, and to
formulate effective field theories that make sense in a restricted
domain.
Decoupling theorems codify the statement that degrees of freedom that
come into play on a high scale do not matter on a low scale.

But the fact that we can formulate a consistent description of
low-energy phenomena without understanding everything that happens
all the way up to very high energies must not blind us to the
additional insights that information from higher energies, or shorter
distances, can bring.  Early in this century, our scientific ancestors
learned that to explain why a table is solid, or why a metal gleams, we
must explore the atomic and molecular structure of matter at a billionth
of human dimensions, where the laws of quantum mechanics take over from
the customs of daily life.  The recent discovery of the top quark in
experiments at a billionth of the atomic scale inspires us to reconsider
how the microworld influences our surroundings.

It is popular to say that top quarks were created in great numbers in the
early moments after the big bang some fifteen billion years ago,
disintegrated in a fraction of a second, and vanished from the scene until
my colleagues learned to create them in the Tevatron at Fermilab.  That
would be reason enough to be interested in top: to learn how it helped sow
the seeds for the primordial universe that has evolved into the world of
diversity and change we live in.  But it is not the whole story; it invests
the top quark with a remoteness that hides its real importance---and
understates the immediacy of particle physics.  The real wonder is that here
and now, every minute of every day, top affects the world around
us.  I would like to close by giving one striking example of top's
influence on the everyday.\r{19}

\begin{figure}[t!]
	\centerline{\BoxedEPSF{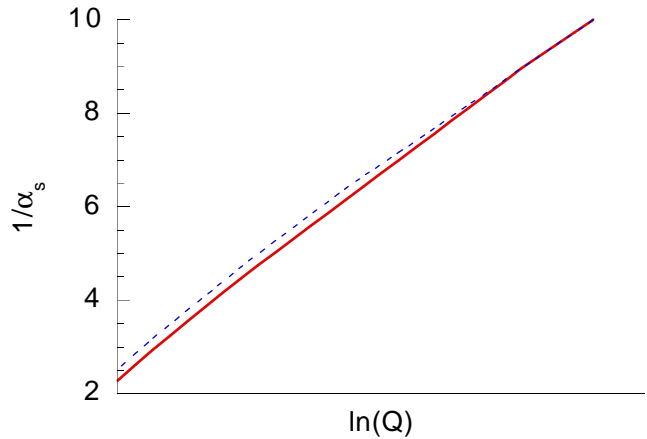 scaled 750}}
	\caption{Two evolutions of the strong coupling constant.}
	\protect\label{fig4}
\end{figure}

Consider a unified theory of the strong, weak, and electromagnetic
in\-ter\-ac\-tions---three-generation $SU(5)$, say---in which all coupling
constants take on a
common value, $\alpha_U$, at some high energy, $M_U$.  If we adopt the
point of view that the value of the coupling constant is fixed at the
unification
scale, then the value of the QCD scale parameter
$\Lambda_{\hbox{{\footnotesize QCD}}}$ depends on the mass of the top
quark.
If we evolve the $SU(3)_c$ coupling, $\alpha_s$, down from the
unification scale in the spirit of Georgi, Quinn, and Weinberg,\r{8} then
the leading-logarithmic behavior is given by
\begin{equation}
1/\alpha_s(Q) = 1/\alpha_U + \frac{21}{6\pi}\ln(Q/M_U)\;\; ,
\end{equation} for $M_U>Q>m_t$.  In the interval between $m_t$ and $m_b$, the
slope $(33-2n_{\!f})/6\pi$ (where $n_{\!f}$ is the number of active quark
flavors) steepens to $23/6\pi$, and then increases by
another $2/6\pi$ at every quark threshold.  At the boundary $Q=Q_n$
between effective field theories with $n-1$ and $n$ active flavors, the
coupling constants $\alpha_s^{(n-1)}(Q_n)$ and $\alpha_s^{(n)}(Q_n)$ must
match.  This behavior is
shown by the solid line in Figure \ref{fig4}.

To discover the dependence of $\Lambda_{\hbox{{\footnotesize QCD}}}$ upon the
top-quark
mass, we use the one-loop evolution equation to calculate $\alpha_s(m_t)$
starting from low energies and from the unification scale, and match:
\begin{equation}
1/\alpha_U  +  {\displaystyle \frac{21}{6\pi}}\ln(m_t/M_U) =
 1/\alpha_s(m_c) - {\displaystyle \frac{25}{6\pi}}\ln(m_c/m_b)
 -{\displaystyle \frac{23}{6\pi}}\ln(m_b/m_t)  \;\; .
 \end{equation} Identifying
 \begin{equation}
 1/\alpha_s(m_c) \equiv {\displaystyle\frac{
 27}{6\pi}}\ln(m_c/\Lambda_{\hbox{{\footnotesize QCD}}})\;\;  ,
\end{equation}  we find that
\begin{equation}
	\Lambda_{\hbox{{\footnotesize QCD}}}=e^{\displaystyle -6\pi/27\alpha_U}
	\left(\frac{M_U}{1 \gev}\right)^{21/27}
	\left(\frac{m_t m_b m_c}{1\gev^{3}}\right)^{2/27}\gev \;\; .
	\label{blank}
\end{equation}

The scale parameter $\Lambda_{\hbox{\footnotesize QCD}}$ is the
only dimensionful parameter in QCD; it determines the scale of the
confinement energy that is the dominant contribution to the proton mass.
We conclude that, in a simple unified theory,
\begin{equation}
	M_{\rm proton} \propto m_t^{2/27} \;\; .
	\label{amazing}
\end{equation}

The dotted line in Figure \ref{fig4} shows how the evolution of
$1/\alpha_s$ changes if the top-quark mass is reduced.
We see from Equations (\ref{blank}) and (\ref{amazing}) that
a factor-of-ten decrease in the top-quark mass would result in a 20\%
decrease in the proton mass.  We can't fully understand the
origin of one of the most important parameters in the everyday
world---the mass of the proton---without
knowing the properties of the top quark.

\section*{Acknowledgements}
It is a pleasure to thank Jean and Kim Tr\^{a}n Thanh V\^{a}n for their
peerless
hospitality and for the boundless energy they invest in the cause of
science, culture, and human understanding.  I salute the organizers of
the Second {\it Rencontres du Vietnam\/} for a pleasant, stimulating, and
exhausting week in Saigon.  I thank Greg Anderson and Gustavo Burdman
for comments on the manuscript.  Fermilab is operated by Universities Research
Association, Inc., under contract DE-AC02-76CHO3000 with the U.S.
Department of Energy.

\def\pl#1#2#3{
       	{\it Phys. Lett. }{\bf #1}, #2 (19#3)}
\def\prl#1#2#3{
	{\it Phys. Rev. Lett. }{\bf #1}, #2 (19#3)}
\def\rmp#1#2#3{
	{\it Rev. Mod. Phys. }{\bf #1}, #2 (19#3)}
\def\prep#1#2#3{
	{\it Phys. Rep. }{\bf #1}, #2 (19#3)}
\def\pr#1#2#3{
	{\it Phys. Rev. D\/}{\bf #1}, #2 (19#3)}
\def\np#1#2#3{
	{\it Nucl. Phys. }{\bf #1}, #2 (19#3)}
\def\zp#1#2#3{
	{\it Z.~Phys. C }{\bf#1}, #2 (19#3)}
\def\apj#1#2#3{
	{\it Astrophys. J. }{\bf #1}, #2 (19#3)}
\def\apjl#1#2#3{
	{\it Astrophys. J. Lett. }{\bf #1}, #2 (19#3)}
\def\ib#1#2#3{
	{\it ibid. }{\bf #1}, #2 (19#3)}
\def\nat#1#2#3{
	{\it Nature (London) }{\bf #1}, #2 (19#3)}
\def\app#1#2#3{
	{\it Acta Phys.~Pol. B\/}{\bf #1}, #2 (19#3)}
\def\ap#1#2#3{
      {\it Ann. Phys. (NY) }{\bf #1}, #2 (19#3)}
\newcommand{\hepph}[1]{(electronic archive: hep--ph/#1)}
\newcommand{\arnps}[3]{{\em Annu. Rev. Nucl. Part. Phys.\/} {\bf#1},
#2 (19#3)}

\begin{reflist}
\single
\frenchspacing

\item These statements are presumptions, not experimental
observations, for the top quark.  It is interesting to ask what
limits, direct and indirect, can be set on the size of top.

\item The distinctions between Dirac and Majorana masses are
elaborated by B. Kayser, These Proceedings; S. Bilenky, These
Proceedings.

\item This characterization is due to E. Eichten, K. Lane, and M.
Peskin, \prl{50}{811}{83}.  For illustrations of the consequences,
see E. Eichten, I. Hinchliffe, K. Lane, and C. Quigg,
\rmp{56}{579}{84}; \ib{58}{1065E}{86}.

\item I. Bars and I. Hinchliffe, \pr{33}{704}{86}.

\item For a review, see H. Harari, ``Composite Quarks and Leptons,''
in \textit{Fundamental Forces,} Proceedings of the Twenty-Seventh
Scottish Universities Summer School in Physics, St. Andrews, 1984,
edited by D. Frame and K. Peach (SUSSP Publications, Edinburgh, 1985),
p.~357.

\item The derivation appears in \S 3.3 of C. Quigg, \textit{Gauge Theories of
the Strong, Weak, and Electromagnetic Interactions} (Addison-Wesley,
Reading, Massachusetts, 1983).

\item H. Georgi and S. L. Glashow, \prl{32}{438}{74}.

\item H. Georgi, H. R. Quinn, and S. Weinberg, \prl{33}{451}{74}.

\item This was emphasized by U. Amaldi, W. de Boer, and H.
F\"{u}rstenau, \pl{B260}{447}{91}.  An equivalent remark is that the
$SU(5)$ prediction for $x_{W}(M_{Z}^{2})$ misses the mark.

\item Peter B. Renton, ``Review of Experimental Results on Precision
Tests of Electroweak Theories,'' 17th International Symposium on
Lepton-Photon Interactions, Beijing, 10--15 August 1995.

\item See K. Nakamura, ``Note on Nucleon Decay,'' in the 1994 Review
of Particle Properties, \pr{50}{1173}{94}; P. Langacker, ``Proton
Decay,'' in \textit{In Celebration of the Discovery of the Neutrino,}
edited by C. E. Lane and R. I. Steinberg (World Scientific, Singapore,
1993), p.~129 \hepph{9210238}.

\item The strategy for deducing the theory of fermion masses at the
unification scale is explained in G. Anderson, \etal,
\pr{49}{3660}{94}.  See also S. Raby, ``Introduction to Theories of
Fermion Masses,'' Ohio State preprint OHSTPY--HEP--T--95--024
\hepph{9501349}.

\item For status reports on the accord between the electroweak theory
and experiment, see the reports by J. Mnich and G. Altarelli, These
Proceedings.

\item Such hints are nevertheless eagerly anticipated, and will be
gratefully received.

\item M. Veltman, \app{12}{437}{81}; C. H. Llewellyn Smith,
\prep{105}{53}{84}.

\item For recent reviews, see J. A. Bagger, ``The Status of
Supersymmetry,'' Johns Hopkins preprint JHU--TIPAC--95021
\hepph{9508392}; X. Tata, ``Supersymmetry: Where it is and how to find
it,'' 1995 TASI Lectures, Hawaii preprint UH--511--833--95
\hepph{9510287}.

\item For a recent historical review, see K. Lane, ``Technicolor,''
Boston University preprint BUHEP--94--26 \hepph{9501249}.

\item For a recent review, see R. S. Chivukula, M. J. Dugan, M.
Golden, and E. H. Simmons, ``Theory of Strongly Interaction
Electroweak Sector,'' \arnps{45}{255}{95} \hepph{9503230}.

\item I owe these insights to discussions with Bob Cahn.  For
additional examples of the links between particle physics and the
quotidian, see R. N. Cahn, ``The Eighteen Parameters of the Standard
Model in Your Everyday Life,'' to appear as a Colloquium in \textit{Reviews
of Modern Physics}.

\end{reflist}

\end{document}